# PAST: A multimodal single-cell foundation model for histopathology and spatial transcriptomics in cancer

Changchun Yang[1,2,3*], Haoyang Li[1,2,3*], Yushuai Wu[4,5], Yilan Zhang[1,2,3], Yifeng Jiao[4,5], Yu Zhang[4,5], Rihan Huang[1,2,3], Yuan Cheng[4,5,6#], Yuan Qi[4,5,6#], Xin Guo[4,5#] and Xin Gao[1,2,3,4#]

[1]Computer Science Program, Computer, Electrical and Mathematical Sciences and Engineering Division, King Abdullah University of Science and Technology (KAUST), Thuwal 23955-6900, Kingdom of Saudi Arabia
[2]Center of Excellence for Smart Health (KCSH), King Abdullah University of Science and Technology (KAUST), Thuwal 23955-6900, Kingdom of Saudi Arabia
[3]Center of Excellence on Generative AI, King Abdullah University of Science and Technology (KAUST), Thuwal 23955-6900, Kingdom of Saudi Arabia
[4]Shanghai Academy of Artificial Intelligence for Science
[5]Artificial Intelligence Innovation and Incubation Institute, Fudan University
[6]Zhongshan Hospital, Fudan University
*These authors contributed equally to this work.
#All correspondence should be addressed to xin.gao@kaust.edu.sa, guoxin@sais.com.cn, qiyuan@fudan.edu.cn, cheng_yuan@fudan.edu.cn.

## Abstract:

While pathology foundation models have transformed cancer image analysis, they often lack integration with molecular data at single-cell resolution, limiting their utility for precision oncology. Here, we present PAST, a pan-cancer single-cell foundation model trained on 20 million paired histopathology images and single-cell transcriptomes spanning multiple tumor types and tissue contexts. By jointly encoding cellular morphology and gene expression, PAST learns unified cross-modal representations that capture both spatial and molecular heterogeneity at the cellular level. This approach enables accurate prediction of single-cell gene expression, virtual molecular staining, and multimodal survival analysis directly from routine pathology slides. Across diverse cancers and downstream tasks, PAST consistently exceeds the performance of existing approaches, demonstrating robust generalizability and scalability. Our work establishes a new paradigm for pathology foundation models, providing a versatile tool for high-resolution spatial omics, mechanistic discovery, and precision cancer research.

## Main

Foundation models in computational pathology have markedly advanced cancer research by leveraging large-scale histopathological image datasets to automate tumor diagnosis, grading, and prognostication [1–5]. These models, trained on millions of whole-slide images (WSIs) across cancer types, extract high-level morphological features that correlate with clinical outcomes, and have demonstrated strong generalization across diverse cohorts [1,2,4,5]. However, despite their success, traditional pathology foundation models are fundamentally limited by their reliance on morphological information alone [4-8], without incorporating fine-grained molecular signatures that are crucial for understanding tumor heterogeneity and guiding precision oncology. These models, while effective at capturing tissue architecture and morphology, overlook the rich molecular landscape encoded in gene expression profiles at the single-cell level—a dimension increasingly recognized as essential for capturing intratumoral diversity and predicting therapeutic response [9-11].

Molecular profiling, particularly at the single-cell level, has become indispensable for precision oncology. Technologies such as single-cell RNA sequencing (scRNA-seq) and spatial transcriptomics (ST) have revealed the complexity of tumor ecosystems [12], enabling the identification of rare cell types, cellular states, and spatially-resolved gene expression patterns [13,14] that drive disease progression and treatment resistance. By preserving the spatial context of gene expression patterns within histological structures, it will reveal intricate tumor-immune interactions, identify clinically actionable molecular signatures with topographic specificity, and ultimately bridge the long-standing gap between morphological phenotypes and their underlying molecular drivers [15, 16].

Recent advances in spatial omics have enabled important progress in correlating tissue morphology with genomic data [17,18], yet significant limitations persist in current computational integration approaches. Most existing methods remain constrained to either bulk-level analyses that average signals across entire tissue sections or coarse tissue-region integrations that fail to capture critical single-cell variations [17,19,20]. These limitations fundamentally restrict their biological relevance and clinical utility in several key aspects. First, by operating at low resolution, such methods obscure rare but clinically important cell populations - including treatment-resistant tumor subclones or immunomodulatory stromal cells - that drive disease progression and therapeutic response [21]. Second, the loss of single-cell spatial context prevents accurate reconstruction of cellular communication networks and tumor microenvironmental niches that are increasingly recognized as determinants of clinical outcomes [22]. The field consequently lacks generalizable frameworks that can simultaneously maintain single-cell resolution, preserve spatial relationships, and scale across diverse cancer types, which are the limitations that must be addressed to realize the full potential of integrated pathology-omics approaches in precision oncology.

To address these limitations, we present a pan-cancer single-cell foundation model that jointly integrates histopathology and transcriptomics at unprecedented scale and resolution (Fig. 1). Leveraging a curated dataset of 20 million paired single-cell histopathology image patches and corresponding single-cell transcriptomes across diverse cancer types, our model (PAST) learns unified representations that capture both morphological and molecular heterogeneity at the cellular level. We comprehensively evaluate PAST across three clinically relevant downstream tasks: (1) single-cell gene-expression prediction from histopathology images, enabling high-resolution spatial transcriptomic inference, whose dense expression maps can be directly used for other single-cell analysis, such as automatic cell-type annotation and discovery of rare subpopulations without additional wet-lab cost; (2) single-cell virtual immunohistochemistry, reconstructing virtual molecular stains from H&E slides, thereby providing interpretable and quantitative read-outs—such as cell-resolved H-scores for ER/PR or HER2 in breast cancer—that can be seamlessly integrated into existing pathology workflows; and (3) multimodal survival prediction, integrating morpho-molecular features for robust outcome prediction, offering mechanistic insights that go beyond the prevailing paradigm of merely enlarging datasets or model capacity and instead highlighting the additive value of cross-modal biology. Together, our approach establishes a scalable and versatile framework for high-resolution spatial omics analysis and precision oncology, unlocking routine, slide-native access to single-cell molecular information and charting a clear path toward clinic-ready, morpho-molecular decision support.

# Results

# Workflow and datasets

We curated a large-scale, pan-cancer dataset by collecting single-cell resolution spatial transcriptomic data and matching histopathology images from 15 cancer types using the Xenium platform [23] (Fig. 1a). For each sample, single cells were spatially mapped to their precise locations on H&E-stained tissue sections, allowing the extraction of image patches centered on each cell. This process enabled us to generate 20 million highly accurate, cell-level image–transcriptome pairs, each capturing both the cellular morphology and its corresponding gene expression profile. The dataset not only spans a wide range of tumor types and tissue microenvironments, but also preserves the spatial relationships between cells, providing a unique resource for integrative morpho-molecular analysis at single-cell resolution.

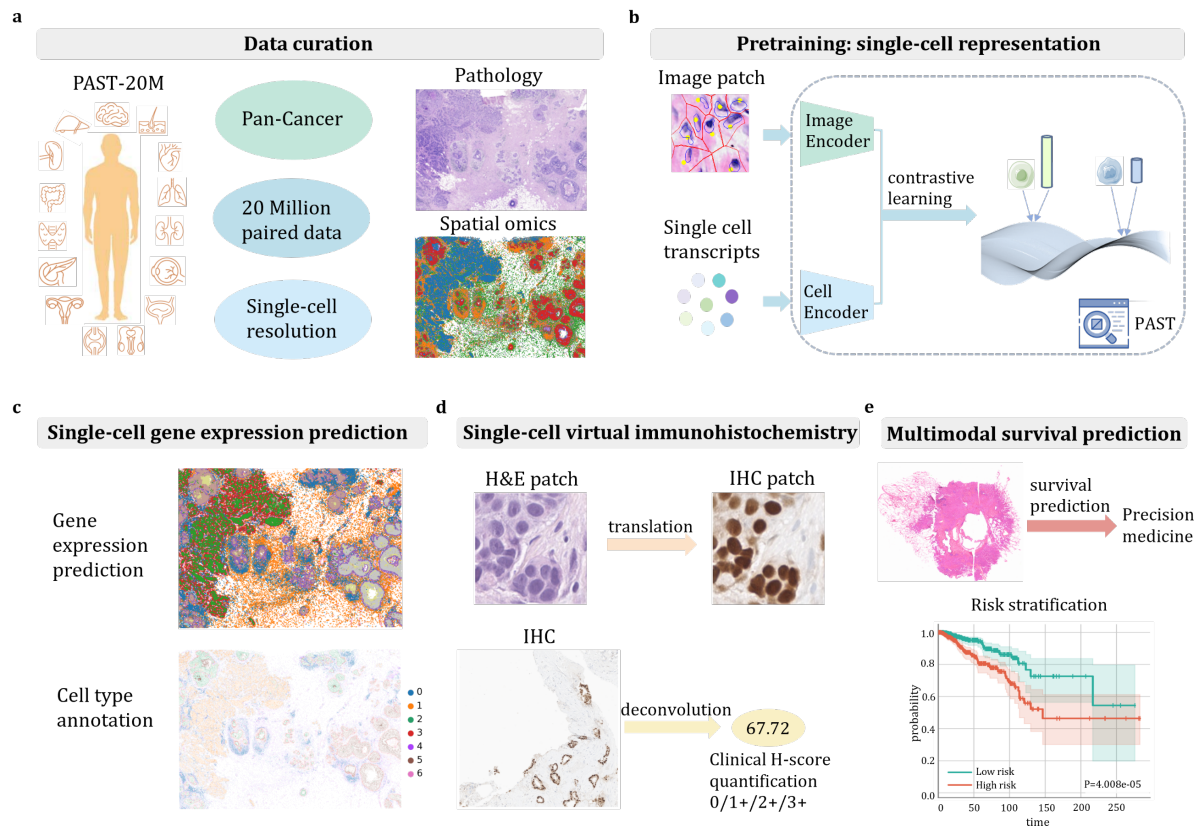

**Fig. 1: PAST: a pan-cancer framework for single-cell integration of pathology and transcriptomics. a** Large-scale curation yields ~20 million paired H&E image patches and matched single-cell transcriptomes spanning 15 distinct tumour types. **b** A dual-encoder backbone is pretrained with contrastive objectives to align image- and RNA-derived embeddings in a shared latent space, thereby coupling each cell's morphology with its molecular state. **c** Single-cell gene-expression prediction: PAST produces slide-wide, single-cell expression maps that can be fed into unsupervised clustering pipelines for automatic cell-type annotation and discovery of rare populations. **d** Single-cell virtual immunohistochemistry: the model renders virtual immunohistochemical or molecular stains at single-cell resolution from routine H&E; downstream DAB de-convolution yields cell-resolved quantitative metrics (for example, H-scores) that are readily interpretable in clinical workflows. **e** Multimodal survival-prediction: patient-level aggregation of joint morpho-molecular features enables superior risk stratification and delivers biological insight beyond image-only or expression-only baselines.

To learn generalizable and biologically meaningful single-cell representations, we developed a contrastive pretraining pipeline based on the CLIP architecture [24] (Fig. 1b). In this framework, each pathology image patch—centered on a target cell—is paired with the gene expression profile of that cell as the positive pair, while profiles of spatially neighboring cells serve as hard negatives. The model consists of dual encoders for image and transcriptomic modalities, which are jointly optimized via a cross-modal contrastive loss. This design enforces the alignment of matched cell-image and gene-expression pairs in the latent space, while pushing apart mismatched pairs. By leveraging both local (cell-centered) and contextual (neighboring cells) information, the model captures subtle morpho-molecular patterns and spatial dependencies, achieving robust pretraining across diverse cancer types and tissue contexts.

Following pretraining, we evaluated the model on three clinically and biologically relevant downstream tasks. First, for single-cell gene expression prediction, we used the pretrained model to infer gene expression levels for individual cells directly from histopathology images, enabling high-resolution spatial transcriptomic reconstruction without additional molecular assays (Fig. 1c). Second, for cell-based virtual immunohistochemistry, we translated H&E image patches to virtual immunohistochemistry (IHC) at the single-cell level, allowing automated and interpretable quantification of clinical HScore, which facilitates objective assessment of biomarker expression (Fig. 1d). Third, for multimodal survival prediction, we integrated the learned morpho-molecular features to predict patient risk, demonstrating improved prognostic performance over conventional pathology foundation models. Across all tasks, our approach highlights the added value of single-cell multi-omics integration, supporting more precise and interpretable cancer diagnosis and prognosis (Fig. 1e).

**Single-cell gene expression prediction**

To assess the ability of PAST to reconstruct single-cell transcriptomes from histological appearance alone, we established a three-tier benchmarking framework of increasing biological complexity, evaluating performance across intra-patient, inter-patient, and cross-cancer generalisation settings (Fig. 2a–c). Pearson correlation between predicted and measured log-normalised gene expression served as the primary evaluation metric. The gene set used for evaluation varied across experimental settings: for intra-patient comparisons, all genes in the patient were included, whereas inter-patient and cross-cancer evaluations employed the union of different genes across training and test cohorts to ensure generalisability across broader transcriptomic space. Full details are provided in the Methods.

When evaluated on held-out cells from the same specimen, PAST achieved a mean correlation of $r = 0.619 \pm 0.099$ (standard deviation across all tumour types), representing substantial gains over ST-Net ($0.447 \pm 0.112$) [20], UNI ($0.536 \pm 0.118$) [1], and CONCH ($0.520 \pm 0.136$) [4] ($P < 0.001$). Strikingly, even when entire patients were withheld during training, PAST retained high accuracy ($r = 0.554 \pm 0.143$), far exceeding the best-performing baseline model (CONCH, $0.477 \pm 0.130$). This generalisation is driven in part by the joint cross-modal alignment introduced during contrastive pretraining, the removal of which led to a marked performance drop (Extended Data Fig. 3). In the most challenging setting—complete exclusion of a tumour type during training—PAST continued to yield robust predictions ($r = 0.319 \pm 0.051$), while all comparator models degraded below 0.25.

In addition to global metrics, visual inspection of predicted spatial expression fields revealed strong agreement with ground-truth ST measurements. For example, PAST accurately reproduced the gland-centric localisation of the luminal marker FOXA1 ($r = 0.76$), while suppressing expression in stromal regions (Fig. 2d), outperforming all baseline models (0.24 – 0.58). These spatially coherent predictions underscore the biological plausibility of the inferred transcriptomes.

We further investigated whether PAST's predictions preserve sufficient granularity to support downstream biological discovery. Clustering of virtual transcriptomes followed by label transfer achieved an adjusted Rand index of 0.53 against original annotations—substantially higher than that achieved by models trained directly on raw ST data (e.g. CONCH, 0.33; Fig. 2e). Importantly, low-abundance yet biologically informative populations were accurately recovered. For instance, ACTA2$^+$ myoepithelial cells (<8 % abundance) were recovered with an F1-score of 0.63, these cells are known to play roles in epithelial integrity and tumour progression, suggesting that PAST captures subtle transcriptomic distinctions reflective of microenvironmental structure and state.

These results position PAST as a versatile foundation model capable of reconstructing high-fidelity spatial transcriptomic profiles from routine histology alone. Its ability to generalise across individuals and cancer types, while preserving fine-grained biological variation, opens new avenues for spatial omics, mechanistic insight, and precision oncology at unprecedented resolution and scale.

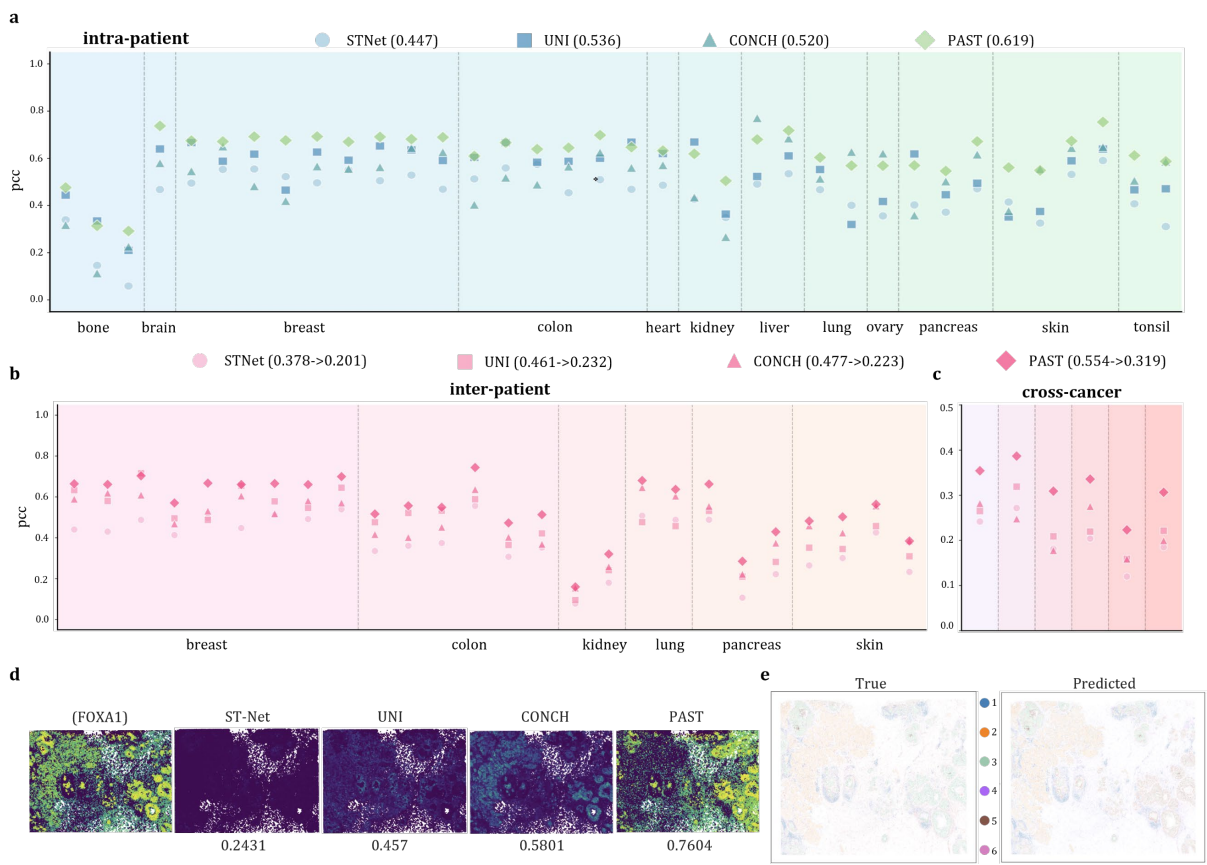

**Fig. 2: PAST accurately reconstructs spatially resolved single-cell gene expression profiles from routine histology.** a Intra-patient evaluation. Performance measured by mean Pearson correlation (r) between predicted and ground-truth log-normalised expression using all genes, with held-out cells from the same specimen (n = 15 tumour types). b Inter-patient, same-cancer evaluation. Models were trained on a subset of patients and tested on unseen individuals within the same cancer type; gene sets correspond to the union of different genes across cohorts. c Cross-cancer generalisation. Models were trained on 14 cancer types and tested on a held-out cancer type, demonstrating robust performance under domain shift. d Spatial expression maps of the luminal marker gene FOXA1 in breast cancer. From left to right: ground-truth ST measurement, ST-Net, UNI, CONCH, and PAST predictions. Slide-level Pearson r values are shown. e Unsupervised cell-type discovery based on PAST-inferred virtual transcriptomes in the cross-cancer setting. Shown are expert annotations (left) and label transfer results obtained by matching predicted transcriptomes to reference profiles and projecting them back into the original spatial context (right).

## Single-cell virtual immunohistochemistry

Immunohistochemistry (IHC) remains the clinical gold standard for measuring protein expression [25] and is routinely used to derive semi-quantitative readouts such as the H-Score [26]. However, the technique is labour-intensive, costly, and often infeasible for large-scale retrospective cohorts due to the unavailability of preserved tissue sections. Existing deep learning approaches typically attempt to predict IHC outcomes directly from H&E-stained slides, treating the task as a coarse slide-level classification problem [27,28]. These methods often overlook critical spatial misalignments between serial sections and fail to capture cellular heterogeneity within the tissue microenvironment.

PAST addresses these limitations through a cell-resolved image-to-image translation module that synthesises virtual IHC stains from standard H&E input (Methods). Operating on the same single-cell image tiles used during pretraining, this module ensures pixel-level registration between the input morphology and the generated stain (Fig. 3a), avoiding artefacts introduced by sectioning offsets between physical H&E and IHC slides. In addition, the shared morpho-molecular latent space learned during foundation-model training serves to regularise the translation process, promoting biologically plausible protein expression patterns and preserving subcellular organisation.

We evaluated the quality of PAST-generated virtual IHC by benchmarking against two representative generative modeling baselines: Improved Diffusion (diffusion-based [35,36]) and RegGAN (GAN-based [37,38]), using the IHC4BC dataset [29], which covers a diverse range of antibody targets. PAST achieved a mean peak signal-to-noise ratio (PSNR) of 25.2 dB and a structural similarity index (SSIM) of 0.61, outperforming both diffusion models (24.3 / 0.57) and GANs (23.8 / 0.60) (Fig. 3b). PSNR reflects PAST's ability to generate higher-quality virtual IHC images with enhanced fidelity, while the SSIM improvement is relatively modest, this is largely due to the motion effects present in many samples, where spatial misalignment between IHC and H&E images occurs. This highlights PAST's unique strength in preserving H&E spatial information during translation (Fig. 3a), ensuring that the generated images retain accurate spatial alignment without distortion. Additionally, PAST demonstrated consistent gains across nuclear markers such as ER and PR, underscoring its capacity to accurately localize subcellular signals and infer protein abundance within nuclei (Extended Data Fig. 4). These results illustrate the robustness and precision of PAST in generating biologically meaningful virtual IHC images, even in the presence of challenging spatial inconsistencies (Fig. 3a).

Visual inspection further confirmed that PAST preserved fine-grained chromatin texture, nuclear shape, and cytoplasmic boundaries, which were often blurred or hallucinated by comparator methods (Fig. 3a). Importantly, cell-wise predictions could be seamlessly aggregated across tissue sections via sliding-window inference to produce continuous virtual IHC slides at whole-slide scale (Fig. 3c), with minimal tiling artefacts and uninterrupted staining gradients. This demonstrates that the approach is computationally tractable and scalable to clinical workflows involving gigapixel images.

To evaluate the quantitative interpretability of virtual stains, we subjected PAST-generated IHC images to a standard clinical pipeline involving nucleus segmentation and DAB deconvolution [45], yielding per-cell optical density values and enabling automated H-Score computation (Fig. 3d). These computational scores exhibited high concordance with expert pathologist assessments (Pearson $r = 0.91$, 95% CI: 0.89–0.93; Fig. 3e), validating the reliability of the predictions in a clinically meaningful context.

These results demonstrate that PAST transforms virtual staining from a black-box aesthetic tool into a biologically faithful and clinically actionable modality. By uniting high-resolution synthesis with transparent, cell-level quantification, PAST enables (i) scalable virtual screening of archival cohorts, (ii) cost-effective companion diagnostics in settings where IHC is inaccessible, and (iii) spatially informed protein biomarker discovery that links subcellular expression patterns to microenvironmental architecture.

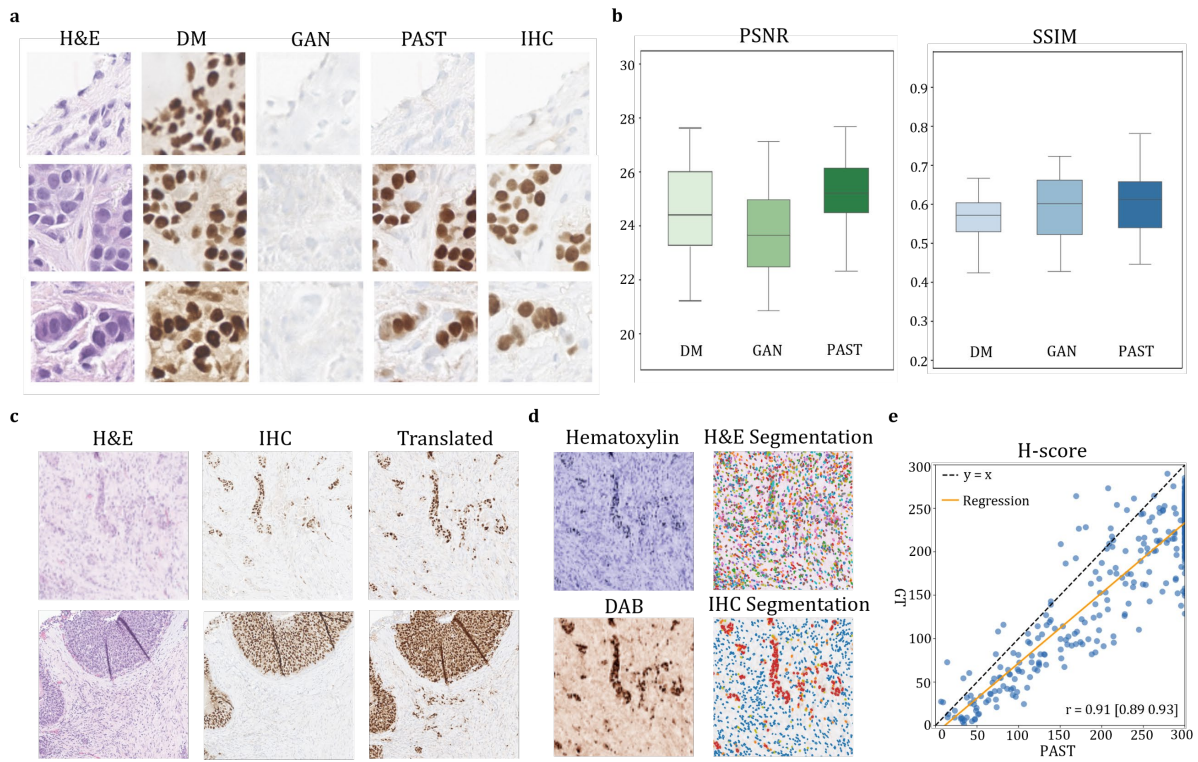

**Fig. 3: PAST generates cell-resolved virtual IHC and enables accurate, interpretable protein quantification.**
**a** Representative single-cell translations from H&E (left) to IHC for a nuclear protein marker. Columns: input H&E, diffusion model (DM), GAN, PAST (ours), and ground-truth IHC. **b** Quantitative comparison of translation fidelity, evaluated by peak signal-to-noise ratio (PSNR; left) and structural similarity index (SSIM; right). **c**, Large-field virtual IHC generated by tiling and stitching cell-wise predictions. Left: input H&E; middle: original IHC; right: PAST prediction. **d**, Computational workflow for automated H-Score estimation from virtual IHC, including nucleus segmentation, DAB deconvolution, and per-cell quantification. **e**, Correlation between PAST-derived H-Scores and expert pathologist annotations. Solid line: identity; inset: Pearson correlation and 95% confidence interval.

## Multimodal survival prediction

Despite the increasing adoption of foundation models in computational pathology, most prognostic frameworks continue to rely solely on morphological information from H&E-stained slides. While model complexity and training data volume have grown substantially, performance gains on survival endpoints often plateau—particularly in datasets with substantial right-censoring. This suggests that improved generalisation may require orthogonal biological signals, not just deeper networks. To this end, we investigated whether the joint morpho-molecular representations learned by PAST could enhance patient-level risk stratification in a clinically realistic setting.

We conducted all experiments within the MCAT [30] multimodal survival framework, which integrates an image encoder with a fixed gene-expression branch. To isolate the impact of image-derived information, we systematically replaced only the image encoder in MCAT across multiple comparator models (ResNet [31], CTransPath [32], UNI [1], PLIP [7], CONCH [4], CHIEF [2], and PAST), keeping the transcriptomic branch unchanged. Beyond this baseline, we introduced PAST-multi, a variant that augments the image input with embeddings from PAST's single-cell–level cell encoder, thereby injecting spatially resolved morpho-molecular context into the multimodal model (Methods).

All models were trained and five-fold cross-validated on six TCGA [33] cancer cohorts (BRCA, LUAD, LUSC, SKCM, COADREAD, and UCEC). Prognostic accuracy was evaluated using two complementary metrics: (i) the standard concordance index (C-index$_1$), which quantifies the consistency between predicted and observed survival times, and (ii) the inverse probability of censoring weighted (IPCW) C-index (C-index$_2$) [34], which adjusts for right-censoring by reweighting observation pairs based on the likelihood of censoring. This censoring-aware variant provides a less biased estimate of model performance, especially in datasets with incomplete follow-up.

Performance varied across cohorts, reflecting tumour-specific morphology–outcome associations. Notably, PAST ranked first in 4 out of 6 datasets, demonstrating its robustness and adaptability across diverse tumour types. Fig. 4a shows results for two datasets, with additional details provided in Extended Data Fig. 5. Substituting PAST into the MCAT framework yielded the highest average C-index$_1$ (0.6953) across all cohorts, outperforming UNI (0.6926) and CONCH (0.6916). Incorporating cell-level embeddings via PAST-multi led to a small decrease in C-index$_1$ (0.6940) but resulted in the highest overall C-index$_2$ (0.636). This indicates that the spatially resolved morpho-molecular features learned by PAST not only capture prognostic information but also mitigate the confounding effects of censoring, yielding more reliable survival predictions in real-world clinical settings.

Kaplan–Meier survival analyses further support the prognostic utility of PAST. On the BRCA cohort, PAST achieved the most pronounced separation between high-risk and low-risk patient groups, with tighter confidence intervals and lower $P$ values (4.03e-2) compared to strong baselines such as UNI (5.52e-1) and CONCH (1.39e-1) (Fig. 4b). This improved stratification indicates that the model not only captures relevant morphological patterns but also leverages molecular context to refine risk predictions.

These findings suggest that biologically informed representation learning can yield meaningful gains in survival prediction. Simple replacement of the image branch with a morphologically superior encoder like PAST improves baseline performance, while further incorporating single-cell molecular context enables robust stratification under clinical censoring. Rather than indiscriminately scaling model size or data quantity, our results argue for principled, biologically grounded model augmentation as a path toward more reliable and generalisable prognostic tools in digital oncology.

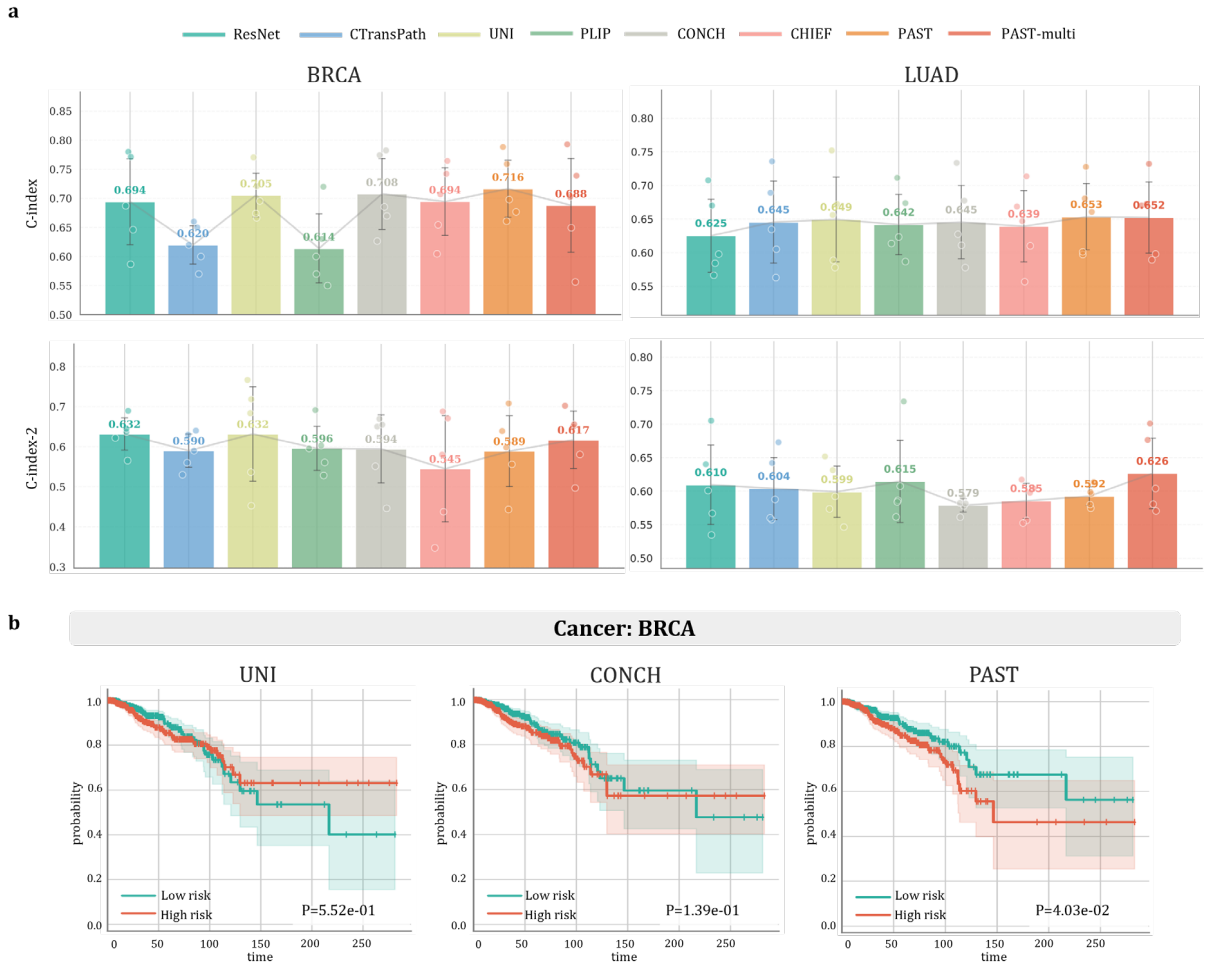

**Fig. 4: Joint morpho-molecular representations improve survival prediction under clinical censoring.**
**a** Prognostic performance of MCAT variants across two representative TCGA cohorts (BRCA, LUAD), evaluated using the standard concordance index (C-index$_1$; top) and its censoring-aware counterpart, the inverse probability of censoring weighted C-index (IPCW C-index, C-index$_2$; bottom). Models differ only in the image encoder: ResNet, CTransPath, UNI, PLIP, CONCH, CHIEF, PAST, and PAST-multi. **b** Kaplan–Meier curves for BRCA stratified by risk scores from UNI, CONCH, and PAST. Shaded areas denote 95% confidence intervals; PAST achieves the clearest separation between risk groups, with the lowest $P$ value and the widest margin between survival curves, indicating enhanced prognostic accuracy and robustness under censoring.

## Discussion

This study establishes a new paradigm for integrative computational pathology by uniting histomorphology and single-cell transcriptomics through a scalable, pan-cancer foundation model. Leveraging over 20 million paired histopathology tiles and single-cell gene expression profiles spanning 15 cancer types, PAST learns biologically grounded representations that generalize across patients, tumor types, and technical platforms. Unlike previous models that rely solely on visual features, PAST captures the molecular underpinnings of tumor biology with single-cell precision, unlocking previously inaccessible spatial insights from routine H&E slides.

To demonstrate the utility of morpho-molecular foundation models, we structured our investigation around three core questions that address the most pressing limitations in digital pathology:

First, *can we accurately predict gene expression at the single-cell level—even under domain shifts across patients and tumor types?* Our results show that PAST consistently outperforms existing models in transcriptome prediction tasks across intra-patient, inter-patient, and cross-cancer settings. This indicates that PAST does not simply memorize training distributions, but instead learns transferable morpho-molecular priors capable of generalizing to unseen biological contexts. The ability to reconstruct spatial transcriptomics from morphology alone paves the way for cost-effective, high-resolution virtual omics in settings where destructive or multiplexed assays are not feasible.

Second, *does our model truly encode morphology at the single-cell level, even in the presence of slight misalignments between H&E and transcriptomic slices?* We address this through a dedicated image-to-image translation task: single-cell virtual immunohistochemistry. By synthesizing IHC stains directly from H&E inputs, PAST achieves higher fidelity and subcellular detail than state-of-the-art generative models, including diffusion and GAN-based methods. The model accurately preserves nuclear texture and cytoplasmic boundaries, while the shared latent space ensures biologically plausible stain patterns. These findings validate that PAST captures interpretable cell-level structure that is robust to minor spatial distortions and staining artifacts.

Third, *can PAST match or exceed the prognostic power of foundation models trained on hundreds of thousands of slides, despite using fewer but more informative multimodal inputs?* Through integration into the MCAT multimodal survival framework, we show that PAST achieves the highest average performance across six TCGA cohorts and delivers the best survival prediction in four out of six individual datasets, particularly excelling in heterogeneous tumor types such as LUAD and COADREAD, demonstrating both robust generalization and cohort-specific utility. Furthermore, the addition of spatially resolved single-cell embeddings (PAST-multi) improves censoring-aware performance, as measured by the inverse probability of censoring weighted (IPCW) concordance index, suggesting enhanced robustness in real-world clinical scenarios. Kaplan–Meier analyses reveal clearer separation of risk groups, emphasizing the prognostic relevance of morpho-molecular context.

Our results show that PAST is not simply a larger or more complex model—it is a biologically informed system designed to answer fundamental questions about how morphology and molecular state co-evolve in cancer. By addressing gene expression imputation, protein marker synthesis, and survival analysis within a unified framework, PAST offers a versatile toolkit for precision oncology and spatial systems biology.

Despite these advances, several limitations remain. Current models are constrained by the limited availability of paired histology and transcriptomic data, particularly at whole-slide resolution, where fewer than 100 examples exist. Expanding such datasets will further improve generalization, especially for rare cancer types and microenvironmental niches. In addition, variability in tissue preparation, staining protocols, and sequencing technologies continues to pose challenges for robust cross-cohort modeling. Improved domain adaptation and multimodal harmonization will be necessary to mitigate these effects.

Looking forward, we anticipate that morpho-molecular foundation models like PAST will form the backbone of future digital pathology pipelines. As more paired datasets become available, and as spatial omics enters the clinical mainstream, these models will be critical for enabling routine histology to serve not only as a visual diagnostic tool but also as a molecular window into tumor biology. By integrating interpretable, spatially resolved molecular information into scalable models, our work charts a path toward more accurate, explainable, and clinically actionable pathology.

# Methods

## Dataset description

### A: Pretraining dataset for the PAST method

For pretraining our PAST model, we curated a large-scale pan-cancer dataset using single-cell resolution spatial transcriptomics data generated by the Xenium platform [23]. Each sample includes a corresponding whole slide image (WSI) stained with H&E and matched single-cell molecular profiles. However, due to technical constraints and tissue handling during sectioning and processing, the spatial coordinates of the transcriptomic data and pathology images are not always perfectly aligned or paired. Therefore, a robust image registration procedure was required to enable high-confidence mapping between modalities at the single-cell level.

We assumed that the transformation between the Xenium transcriptomic domain and the pathology image domain is rigid within local regions. To estimate the coordinate transformation, we manually selected several keypoints on both the transcriptomic coordinate map and the corresponding WSI. For each set of matched keypoints, we computed the rigid transformation matrix $T$ by Kabsch algorithm [39]. To account for tissue heterogeneity and local deformations, we selected multiple sets of keypoints from distinct regions on each WSI, including areas with high or low cell density, tumor-enriched zones, and non-tumoral regions. The resulting transformation matrices from different keypoint sets were then combined and adjusted to maximize registration accuracy across the slide. This enabled us to robustly align single-cell molecular profiles with the corresponding histopathological context.

Cell segmentation masks, including nuclear and cell boundaries, were obtained from DAPI-based segmentation provided by the Xenium platform. Only high-quality, confidently segmented cells were retained for downstream analysis. Using this multi-step registration and quality control approach, we constructed a paired dataset comprising 19,339,608 single-cell image-transcriptome pairs across 15 distinct cancer types (see Extended Data Table 1 for details). This resource forms the foundation for pretraining and evaluating our pan-cancer, cross-modal representation learning framework.

### B: Dataset for downstream tasks

For downstream evaluation and benchmarking, we curated and preprocessed multiple datasets tailored to the specific requirements of each task.

**Single-cell gene expression prediction:** We utilized single-cell spatial transcriptomics datasets generated on the Xenium platform [23], following preprocessing and quality control, we established three distinct generalization settings. Intra-patient: Model training and evaluation were performed on different regions from the same patient. This setting comprised 32 slices from multiple anatomical sites. Inter-patient: The model was evaluated across different patients, testing its ability to generalize to new individuals. Cross-cancer: The most stringent setting, where the model was trained on one or more cancer types and tested on others, assessing cross-tissue generalizability. For the inter-patient and cross-cancer experiments, we selected six cancer types, covering a total of 24 slices. Cancer types with only a single patient were excluded from these generalization analyses to ensure statistical validity.

**Single-cell virtual immunohistochemistry:** For virtual IHC image translation tasks, we used the publicly available IHC4BC (ImmunoHistoChemistry for Breast Cancer) dataset [29]. After removing images with large white borders and those with poorly defined cellular boundaries, we curated training cohorts, resulting in 3,600 images for estrogen receptor (ER) and 3,200 images for progesterone receptor (PR) staining prediction.

**Multimodal survival prediction:** To evaluate the prognostic utility of our multimodal models, we employed six large-scale cancer cohorts from The Cancer Genome Atlas (TCGA): breast invasive carcinoma (BRCA), lung adenocarcinoma (LUAD), lung squamous cell carcinoma (LUSC), skin cutaneous melanoma (SKCM), colorectal adenocarcinoma (COADREAD), and uterine corpus endometrial carcinoma (UCEC). For each dataset, both histopathology slides and paired transcriptomic profiles were used for MCAT framework.

# PAST method

**Data Input Preparation**

**Image Encoder Input Preparation:** For each cell, we utilized the segmentation masks provided by the Xenium platform to extract both the cell and nuclear boundaries. The centroid of each cell $c$ is calculated as follows:

$$x_c = \frac{1}{N_c} \sum_{j=1}^{N_c} p_j$$

where $p_j$ denotes the coordinates of the $j$th pixel within the cell mask, and $N_c$ is the total number of pixels in cell $c$. Using this centroid, we extracted a square image patch of size $L \times L$ (with $L$=224) centered at $x_c$ from the corresponding H&E-stained whole slide image. All image patches were normalized prior to model input using the standard mean and standard deviation (mean = [0.485, 0.456, 0.406], std = [0.229, 0.224, 0.225]) to ensure consistency and comparability across samples.

**Cell Encoder Input Preparation:** To address differences in sequencing depth and batch effects inherent to spatial transcriptomics, and for enhanced generalizability and future extension, we adapted and extended the tokenization strategies from Geneformer [40] and Nicheformer [41]. Rather than directly using absolute gene expression values, we ranked all detected genes within each cell based on normalized expression and used their indices in descending order as tokens. Specifically, for a given cell, let $gex_i$ denote the normalized expression of gene $i$:

$$T = \{idx(gex_{(1)}), idx(gex_{(2)}), \ldots, idx(gex_{(n)}) \mid gex_{(j)} \geq gex_{(j+1)}; gex_{(j)} \neq 0\}$$

where $idx(gex_{(j)})$ maps the $j$th highest expressed gene to its index in a predefined gene vocabulary. The total number of genes is 20,310, consistent with Nicheformer, which indexed 16,981 orthologous genes, 3,178 human-specific genes, and 151 mouse-specific genes—providing a comprehensive and sufficiently large basis for cross-species and cross-study analyses. This ranking-based approach minimizes the influence of batch effects and technical variation compared to using absolute expression values. To further enhance the biological context and model flexibility, we prepended three semantic tokens to each cell input: $<ASSAY>$ (set to "xenium" in this study), $<MODALITY>$ (set to "spatial"), and $<ORGANISM>$ (human or mouse). The final token sequence for a cell $i$ is thus:

$$T^i = \{assay^i, organism^i, modality^i, idx(gex_{(1)}{}^i), idx(gex_{(2)}{}^i), \ldots, idx(gex_{(n)}{}^i)\}$$

where $assay^i$, $organism^i$, and $modality^i$ encode the relevant metadata for that cell. For all experiments, we set the maximum sequence length $N$ to 300, reflecting the average number of genes expressed per cell in the Xenium single-cell spatial transcriptomics dataset (after metadata tokens). This threshold was determined empirically based on the observed distribution of nonzero expressed genes per cell. If a cell contained fewer than $N$ expressed genes, we appended special $<PAD>$ tokens to ensure uniform input length across batches.

**PAST architecture**

Given a batch of $B$ paired image patches $V \in R^{B \times L \times L}$ and corresponding per-cell gene index sequences $G \in N^{B \times N}$, the PAST architecture is designed to jointly encode cellular morphology, gene profile, and local spatial context.

**Image Encoder:** For each image patch, we employ a large Vision Transformer (ViT) [42] as the image encoder $f_{img}$. Each patch of size $L \times L$ is tokenized and processed through multiple transformer layers, yielding a fixed-dimensional embedding ($D = 1024$) for each cell-centric image region. To align representation scales across modalities, image features are further projected into a shared embedding space ($D = 256$) using a learnable projection head.

**Cell encoder with spatial context:** Inspired by Nicheformer, but extending beyond single-cell context, our cell encoder operates in two synergistic branches. Single-cell transformer branch: The gene index sequence for the center cell, constructed as described above, is embedded and processed with a stack of transformer blocks. Each block comprises multi-head self-attention and position-wise feedforward networks:

$$h_{l+1}^i = TransformerBlock_l(h_l^i), h_0^i = Embed(G^i)$$

where $h_l^i \in R^{N \times D}$ is the hidden state at layer $l$, and learnable positional embeddings are used. The final hidden state is pooled to obtain a single embedding ($D = 512$) for the center cell. Spatial graph branch: To capture the spatial microenvironment, for each center cell, we construct a local cell graph incorporating the center and its neighboring cells. Each node is initialized with the gene index embedding processed as above. The graph is encoded using a multi-layer Graph Attention Network (GAT) [43], to aggregate features from the local cell neighborhood:

$$z^i = GAT(\{h_0^j\}_{j \in N(i)}, E)$$

where $N(i)$ denotes the set of neighboring cells, and $E$ is the edge set constructed based on Euclidean distances between the center cell and its surrounding neighbors. The transformer-derived center cell feature $f_{trans}$ and the spatial graph feature $f_{graph}$ are integrated using attention-based fusion strategy. We employ a multi-head attention module, where the transformer feature acts as the query and both the transformer and spatial graph features are concatenated as keys and values:

$$K = [f_{trans}, f_{graph}], V = [f_{trans}, f_{graph}]$$

$$f_{fused} = MultiHeadAttn(Q = f_{trans}, K, V)$$

where $MultiHeadAttn(\cdot)$ denotes the standard multi-head self-attention operation. The fused feature $f_{fused}$ is then projected to obtain the final cell embedding ($D = 256$).

**PAST pretraining**

PAST was pretrained for approximately 12 days using 8 Nvidia A100 80GB GPUs. We used the AdamW optimizer [44] with a batch size of 48. The learning rate was linearly increased from a minimum of $1 \times 10^{-6}$ to a peak of $1 \times 10^{-3}$ during the first epoch (linear warmup), then decayed according to a cosine schedule for the remainder 15 epoches.

Our pretraining objective is based on a contrastive learning approach inspired by CLIP [19, 24], but with an adjusted loss to account for the biological context, where multiple cells with similar gene expression or morphological profiles are likely to appear within the same batch. To prevent the model from pulling apart cells with similar profiles, we incorporate a similarity-adjusted target, which encourages coherence in the learned joint embedding space. Given a batch of $B$ paired image and cell representations, let $H_v \in R^{B \times d}$ denote the normalized image embeddings and $H_g \in R^{B \times d}$ the normalized fused cell embeddings. We first compute pairwise similarities:

$$sim(H_v, H_g) = H_g H_v^T$$

To account for internal similarities within each modality, we also compute:

$$sim(H_v, H_v) = H_v H_v^T, sim(H_g, H_g) = H_g H_g^T$$

A similarity-adjusted target matrix is then defined as:

$$target = softmax(\frac{sim(H_g, H_g) + sim(H_v, H_v)}{2 \cdot \tau})$$

where $\tau$ is a learnable temperature parameter controlling the sharpness of the distribution. The final loss is the mean of cross-entropy (CE) losses in both directions (cell-to-image and image-to-cell):

$$L = mean(CE(sim(H_v, H_g), target) + CE(sim(H_v, H_g)^T, target^T))$$

This training strategy encourages paired image and cell representations to align, while also increasing the coherence of the joint embedding for cells or images with similar biological properties.

## Downstream tasks

**Single-cell gene expression prediction**

To evaluate the capacity of PAST to infer single-cell transcriptomic profiles directly from morphological features, we devised a transfer learning protocol leveraging the pretrained ViT-large backbone. Specifically, the final layer of the image encoder is replaced with a linear head whose output dimensionality matches the number of genes under prediction. This output dimension is dynamically set according to the evaluation context: for intra-patient experiments, all genes detected in the given patient are included; for inter-patient comparisons within a cancer type, the union of all genes detected across patients is used; and for cross-cancer generalisation, the union of all genes detected across the full set of cancer types is adopted. Formally, let $v_i$ denote the image embedding for cell $i$, and let $f_{gene}$ be the trainable linear head. The predicted gene expression vector for cell $i$ is given by:

$$\hat{g}_i = f_{gene}(v_i) \in R^d$$

where $d$ is determined as above for each experimental setting.

Performance is assessed using the Pearson correlation coefficient (PCC) between predicted and ground-truth log-normalised gene expression value:

$$PCC_{gene} = \frac{1}{d}\sum_{j=1}^{d} corr(\hat{g}_{:,j}, g_{:,j})$$

where $\hat{g}_{:,j}$ and $g_{:,j}$ represent the predicted and ground-truth expression values of gene $j$ across all test cells, and $d$ is the number of genes.

To further assess the biological granularity of predicted expression profiles, we performed unsupervised clustering and label transfer. Specifically, both predicted and ground-truth gene expression matrices were independently clustered using the Leiden algorithm. For each cluster, the top 5 marker genes were identified based on differential expression. Cluster correspondence between predicted and actual groupings was established by ranking the PCC between sets of top marker genes:

$$PCC_{cluster} = corr(\{\hat{g}_{i,top5}\}, \{g_{i,top5}\})$$

where $g_{i,top5}$ denotes the vector of top 5 genes for cluster $i$. This enables accurate mapping of predicted clusters to real cell types, supporting downstream cell annotation transfer and biological interpretation.

**Single-cell virtual immunohistochemistry**

To enable high-resolution virtual staining from standard H&E images, we integrated the pretrained PAST image encoder into a generative translation framework. Specifically, the image encoder's feature embedding for each cell tile was fused into the generator network via feature-wise linear modulation (FiLM) conditioning. At each relevant generator stage, the PAST embedding is processed into per-channel scale and bias vectors, which modulate the generator activations as follows:

$$FiLM(x, embedding) = x \times (1 + \gamma) + \beta$$

where $\gamma, \beta$ are derived from the projected PAST embedding. This conditioning enforces cell- and context-specific translation, leveraging the morpho-molecular knowledge learned during pretraining. The image quality of predicted IHC stains was assessed using both peak signal-to-noise ratio (PSNR) and structural similarity index (SSIM), computed between the generated and real IHC images at the cell-tile level:

$$PSNR = 10 \cdot log_{10}(\frac{MAX^2}{MSE})$$

where $MAX$ is the maximum value of the pixel in the image, and

$$MSE = \frac{1}{N}\sum_{i=1}^{N}(x_i - y_i)^2$$

with $x_i$ and $y_i$ denoting the pixel intensities of the predicted and ground-truth images, respectively. The structural similarity index (SSIM) is calculated as

$$SSIM(x, y) = \frac{(2\mu_x\mu_y + C_1)(2\sigma_{xy} + C_2)}{(\mu_x^2 + \mu_y^2 + C_1)(\sigma_x^2 + \sigma_y^2 + C_2)}$$

where $\mu_x$ and $\mu_y$ are the mean pixel intensities of the predicted and ground-truth images, $\sigma_x^2$ and $\sigma_y^2$ are their respective variances, $\sigma_{xy}$ is the covariance between the two images, and $C_1$, $C_2$ are small constants to stabilize the division.

For clinical interpretability, we applied a standard H-DAB analysis pipeline to the virtual IHC images to compute per-cell H-scores and related metrics. Each H&E/IHC pair (3000 × 3000 px) underwent the following process: Nuclei segmentation: StarDist [46] was used to segment nuclei, with segmentation performed on the H&E image for reliable nucleus counting. Color deconvolution [45]: The brown DAB channel was extracted from the IHC image using conventional color deconvolution. Per-nucleus quantification: For each nucleus, the average DAB optical density was measured, and thresholds [29] were used to classify nuclei as weakly, moderately, or strongly stained. H-score computation:

$$H - score = (\% \, weak \, nuclei \times 100) + (\% \, moderate \times 200) + (\% \, strong \times 300)$$

The denominator for percentages was the total number of nuclei in the corresponding H&E patch.

## Multimodal survival prediction

To evaluate the clinical relevance of PAST's morpho-molecular representations, we conducted patient-level survival prediction using the MCAT [30] multimodal framework. MCAT integrates image-derived features and genomic attributes by embedding gene categories into a structured latent space. For all experiments and baselines, we adopted the same MCAT architecture for fair comparison (only replce the image encoder part). For the PAST-multi variant, our aim was not to modify the underlying MCAT framework, but rather to probe the added value of PAST's single-cell information. Specifically, we treated the output of PAST's cell encoder as an additional "genomic" channel: for each patient, the MCAT genomic branch input was augmented with the PAST-derived embedding, effectively extending the set of gene-category features from $\{B_n\}_{n=1}^{N}$ to $\{B_n\}_{n=1}^{N+1}$. This allowed direct comparison to the original MCAT formulation.

Consistent with prior work, we systematically replaced the image encoder in MCAT with various backbone models (ResNet [31], CTransPath [32], UNI [1], PLIP [7], CONCH [4], CHIEF [2], and PAST), without altering the transcriptomic branch. All models were trained and evaluated via five-fold cross-validation on six TCGA [33] cancer cohorts (BRCA, LUAD, LUSC, SKCM, COADREAD, UCEC).

Prognostic performance was assessed using the standard concordance index (C-index[1]), which measures the agreement between predicted and observed survival times,

$$C - index_1 = \frac{1}{|\mathcal{P}|} \sum_{(i,j) \in \mathcal{P}} I((\hat{r}_i > \hat{r}_j) \wedge (t_i > t_j))$$

where $\mathcal{P}$ is the set of all comparable (uncensored) patient pairs, $\hat{r}_i$ and $\hat{r}_j$ are the predicted risk scores, $t_i$ and $t_j$ are the observed survival times, and $I(\cdot)$ is the indicator function. and To account for right-censoring, the IPCW C-index (C-index[2]) [34] weighs each comparable pair by the inverse of the probability that both individuals are uncensored at the relevant time point:

$$C - index_2 = \frac{1}{\sum_{(i,j)\in\mathcal{P}} w_{ij}} \sum_{(i,j)\in\mathcal{P}} w_{ij} \cdot I((\hat{r}_i > \hat{r}_j) \wedge (t_i > t_j))$$

where $w_{ij} = \frac{\delta_j}{\hat{G}(t_j)}$, $\delta_j$ is the event indicator (1 if event observed, 0 if censored), and $\hat{G}(t_j)$ is the Kaplan–Meier estimate of the probability of being uncensored at time $t_j$. Kaplan–Meier survival analysis was performed to visualize risk group stratification, with log-rank tests used to assess statistical significance.

## Data Availability

Public datasets for pretraining and single-cell gene expression prediction (Xenium, https://www.10xgenomics.com/datasets?query=&page=1&configure%5BhitsPerPage%5D=50&configure%5BmaxValuesPerFacet%5D=1000&refinementList%5Bproduct.name%5D%5B0%5D=In%20Situ%20Gene%20Expression&refinementList%5Bplatform%5D%5B0%5D=Xenium%20In%20Situ) [23], single-cell virtual immunohistochemistry (IHC4BC, https://ihc4bc.github.io/) [29], multimodal survival prediction (https://portal.gdc.cancer.gov/projects/TCGA-BRCA, https://portal.gdc.cancer.gov/projects/tcga-luad, https://portal.gdc.cancer.gov/projects/tcga-lusc, https://portal.gdc.cancer.gov/projects/TCGA-SKCM, https://portal.gdc.cancer.gov/projects/TCGA-COAD, https://portal.gdc.cancer.gov/projects/TCGA-UCEC) are publicly available from their original links. The authors declare that all other data supporting the findings of this study are available within the paper and its supplementary information files.

## Code Availability

The source code and the trained models for a working version of PAST is available at https://github.com/Changchun-Yang/past_sc.